\documentclass[journal]{IEEEtran}

\ifCLASSINFOpdf
\else
\fi

\usepackage{amsmath,amsfonts}
\usepackage{amsthm}
\usepackage{algorithmic}
\usepackage{verbatim}
\usepackage{graphicx}
\usepackage{cite}
\usepackage[ruled,linesnumbered]{algorithm2e}
\usepackage{cancel}
\usepackage{indentfirst}
\usepackage{epstopdf}
\usepackage{caption}
\usepackage{color}
\usepackage{subfigure}
\usepackage[colorlinks,citecolor=blue,linkcolor=blue,urlcolor=blue]{hyperref}
\usepackage{cite}

\begin{document}

\title{\LARGE  From Partial Calibration to Full Potential: A Two-Stage 	\\ Sparse DOA Estimation  for Incoherently-Distributed Sources with Gain-Phase Uncertainty   }
\author{He Xu, Tuo Wu, Wei Liu,  Maged Elkashlan,  Naofal Al-Dhahir, \emph{Fellow, IEEE},\\ M\'{e}rouane Debbah, \emph{Fellow, IEEE}, Chau Yuen, \emph{Fellow, IEEE}, and Hing Cheung So, \emph{Fellow, IEEE} 
\thanks{ (\textit{Corresponding author: Tuo Wu.})
	
H. Xu is with the School of Cyber Science and Engineering, Ningbo University of Technology, Ningbo 315211, China (E-mail: $\rm xuhebest@sina.com$).
T. Wu and C. Yuen are with the School of Electrical and Electronic Engineering, Nanyang Technological University 639798, Singapore (E-mail: $\rm \{tuo.wu, chau.yuen\}@ntu.edu.sg$).
W. Liu is with the Department of Electrical and Electronic Engineering, Hong Kong Polytechnic University, Kowloon, Hong Kong (E-mail: $\rm wliu.eee@gmail.com$).
M.~Elkashlan is with the School of Electronic Engineering and Computer Science at Queen Mary University of London, London E1 4NS, U.K. (E-mail: $\rm maged.elkashlan@qmul.ac.uk$).
Naofal Al-Dhahir is with the Department of Electrical and Computer Engineering, The University of Texas at Dallas, Richardson, TX 75080 USA (E-mail: $\rm aldhahir@utdallas.edu$).
M. Debbah is with  KU 6G Research Center, Department of Computer and Information Engineering, Khalifa University, Abu Dhabi 127788, UAE (E-mail: $\rm merouane.debbah@ku.ac.ae$).
H. C. So is with the Department of Electrical Engineering City University of Hong Kong, Hong Kong, China. (E-mail: $\rm hcso@ee.cityu.edu.hk$).
}}
\markboth{IEEE Internet of Things Journal,~Vol.~XX, No.~XX, XX~2025}
{Shell \MakeLowercase{\textit{et al.}}: A Sample Article Using IEEEtran.cls for IEEE Journals}
\maketitle

\begin{abstract}
Direction-of-arrival (DOA) estimation for incoherently distributed (ID) sources is essential in multipath wireless communication scenarios, yet it remains challenging due to the combined effects of angular spread and gain-phase uncertainties in antenna arrays. This paper presents a two-stage sparse DOA estimation framework, transitioning from \textit{partial calibration to full potential}, under the generalized array manifold (GAM) framework. In the first stage, coarse DOA estimates are obtained by exploiting the output from a subset of partly-calibrated arrays (PCAs). In the second stage, these estimates are utilized to determine and compensate for gain-phase uncertainties across all array elements. Then a sparse total least-squares optimization problem is  formulated and solved via   alternating descent   to refine the DOA estimates. Simulation results demonstrate that the proposed method attained improved estimation accuracy compared to existing approaches, while maintaining robustness against both noise and angular spread effects in practical multipath environments.
\end{abstract}


\IEEEpeerreviewmaketitle
 
\section{Introduction}
\IEEEPARstart{D}{irection}-of-arrival (DOA) estimation remains a foundational task in internet of vehicles (IoVs), supporting functions such as channel estimation and vehicle positioning in multi-path environments \cite{ref1,ref2}. Among the two major models, namely, coherently-distributed (CD) and incoherently-distributed (ID) sources \cite{ref3}, the latter modeling approaches are more common in real cellular networks with rich scattering, making accurate DOA estimation vital for robust system performance.

Over the years, numerous ID source estimation methods have been proposed, including classical techniques such as the covariance matching estimator (COMET) \cite{ref4}, ESPRIT-based method \cite{ref5}, and beamspace transformation-based estimator (BTBET) \cite{ref6}. Additionally, rank-reduction (RARE) estimator \cite{ref7} and its two-stage variant \cite{ref8}, are able to handle  mixed far-field and near-field ID sources. However, these subspace-based approaches typically require prior knowledge of the number of sources,  which may not be available under low signal-to-noise ratio (SNR) conditions. Moreover, subspace methods often  suffer performance degradation when the signal subspace dimension is uncertain or when noise levels are high, highlighting the need for more resilient alternatives. Another significant issue not adequately addressed by these methods is the gain-phase uncertainty inherent in practical radio frequency (RF) chains \cite{ref9}. Both theoretical analysis and algorithms investigations confirm that these uncertainties can severely degrade estimation accuracy \cite{ref10,ref11,ref12}. Although recent studies tackled this challenge through partly-calibrated arrays (PCAs) \cite{ref13} and augmented generalized array manifold (GAM) approaches \cite{ref14}, most solutions focus on isolated aspects, such as merely compensating gain-phase errors, while leaving other critical factors, especially the angular spread in ID sources caused by multipath propagation, unaddressed in a unified framework.

To overcome these shortcomings, sparse reconstruction provides a promising solution. Unlike subspace-based algorithms, sparsity-driven approaches do not hinge on prior knowledge of the number of sources, thereby achieving better performance in low-SNR regimes. Besides, sparse methods inherently accommodate practical imperfections, such as gain-phase uncertainties, making them well-suited for the combined challenges posed by ID sources and multipath propagation. To this end, we propose a two-stage DOA estimator  for ID sources that fully capitalizes on the sparsity framework and leverages PCAs, while adopting  GAM to accurately capture the small angular deviations characteristic of ID sources.  Our approach proceeds in two stages.  First, a set of  consecutive calibrated sensors is used to extract the first column of the sample covariance matrix, and  we form an augmented covariance vector to simplify the initial estimation while mitigating noise effects and partial angular spread interference. Subsequently, the coarse DOA estimates enable us to identify and compensate for gain-phase uncertainties in the uncalibrated sensors, allowing a refined covariance structure to be constructed using all array elements. This refined structure is then processed via a sparse total least-squares minimization, further enhancing DOA accuracy. Simulation results show that this two-stage strategy outperforms competing  subspace estimators across a broad range of SNR levels.

	The remainder of this article is as follows: the problem formulation is given in Section II. The proposed two-stage DOA estimator is presented in detail in Section III. Numerical   results  are provided in Section IV. Conclusions are drawn in Section V.
 
\section{Problem Formulation}
Consider $K$ ID sources impinging on an $M$-element uniform linear array (ULA) with half wavelength inter-element spacing, and each ID source signal propagates along $L$ distinct paths. Suppose that there are $M_c$ consecutive calibrated sensors, then the array output after considering the gain-phase uncertainties at time instant $t$ can be expressed as \cite{ref7,ref8}
\begin{equation}\label{1}
{\bf{z}}\left( t \right) = {\bf{\Xi }}\left( {\bf{g}} \right)\sum\limits_{k = 1}^K {{s_k}\left( t \right)} \sum\limits_{l = 1}^L {{\gamma _{k,l}}\left( t \right)} {\bf{a}}\left( {{\theta _{k,l}}(t)} \right) + {\bf{e}}\left( t \right),
\end{equation}
where ${\bf{\Xi }}\left( {\bf{g}} \right) = {\mathop{\rm diag}\nolimits} \left( {\bf{g}} \right)$ stands for the gain-phase uncertainty matrix with ${\bf{g}} = {\left[ {{{\bf{1}}_{1 \times Mc}},{g_{M_c + 1}}, \ldots ,{g_M}} \right]^T}$ with ${g_{m}} = {\rho _{m}}{e^{j{\phi _{m}}}}$, $m \in \left[ {M_c + 1,M} \right]$, while ${\rho _{m}}$ and $\phi _{m}$ represent the gain and phase uncertainties, respectively. $\left\{ {{s_k}\left( t \right)} \right\}_{k = 1}^K$ are $K$ far-field narrowband uncorrelated ID signals and ${\bf{e}}\left( t \right)$ is the additive Gaussian white noise. ${{\gamma _{k,l}}\left( t \right)}$ is the complex path gain of the $l$-th path associated with the $k$-th ID source, which is zero-mean and temporally independent and identically distributed (i.i.d.) random variable with unknown variance $\sigma _{{\gamma _k}}^2/L$. ${\bf{a}}\left( {{\theta _{k,l}}(t)} \right)$ represents the array steering vector of ${\theta _{k,l}}(t)$, with ${\theta _{k,l}}\left( t \right) = {\theta _k} + {{\tilde \theta }_{k,l}}\left( t \right)$, ${\theta _k}$ is the DOA of the $k$-th signal, and ${{\tilde \theta }_{k,l}}\left( t \right)$ is the real-valued angular deviation, which normally follows Gaussian or uniform distribution with unknown variance $\sigma _{{\theta _k}}^2$ \cite{ref7, ref15}.

To simplify the mathematical modeling of multipath propagation in array signal processing, the GAM model \footnote{The GAM model describes multipath propagation by combining the main signal direction with small angle deviations due to scattering, allowing for effective mathematical treatment while capturing the essential multipath characteristics.} with a small $\sigma_{{\theta_k}}^2$ is adopted, where ${\bf{a}}\left( {{\theta_{k,l}}(t)} \right)$ is approximated as:
\begin{equation}\label{2}
{\bf{a}}\left( {{\theta _{k,l}}(t)} \right) \approx {\bf{a}}\left( {{\theta _k}} \right) + {\bf{a'}}\left( {{\theta _k}} \right){{\tilde \theta }_{k,l}}\left( t \right),
\end{equation}
where ${\bf{a'}}\left( {{\theta _k}} \right) = {{\partial {\bf{a}}\left( {{\theta _k}} \right)} \mathord{\left/
 {\vphantom {{\partial {\bf{a}}\left( {{\theta _k}} \right)} {{\theta _k}}}} \right.
 \kern-\nulldelimiterspace} {{\theta _k}}}$ and
\begin{equation}\label{3}
{\bf{a}}\left( {{\theta _k}} \right) = {\left[ {1,{e^{ - j\pi \sin {\theta _k}}}, \ldots ,{e^{ - j\left( {M - 1} \right)\pi \sin {\theta _k}}}} \right]^T}.
\end{equation}

Consequently, ${\bf{z}}\left( t \right)$ is expressed as
\begin{multline}\label{4}
{\bf{z}}\left( t \right) \approx {\bf{\Xi }}\left( {\bf{g}} \right)\left( {\sum\limits_{k = 1}^K {{\bf{a}}\left( {{\theta _k}} \right){{\bar s}_k}\left( t \right)}  + \sum\limits_{k = 1}^K {{\bf{a'}}\left( {{\theta _k}} \right){{\tilde s}_k}\left( t \right)} } \right) + {\bf{e}}\left( t \right) \\
= {\bf{\Xi }}\left( {\bf{g}} \right)\left( {{\bf{As}}\left( t \right) + {\bf{A'\tilde s}}\left( t \right)} \right) + {\bf{e}}\left( t \right),\quad\quad\quad\quad\quad\quad
\end{multline}
where ${\bf{A}}= \left[ {{\bf{a}}\left( {{\theta _1}} \right), \ldots ,{\bf{a}}\left( {{\theta _K}} \right)} \right]$, ${\bf{A'}} = \left[ {{\bf{a'}}\left( {{\theta _1}} \right), \ldots ,{\bf{a'}}\left( {{\theta _K}} \right)} \right]$, ${\bf{s}}\left( t \right) = {\left[ {{s_1}\left( t \right),\ldots ,{s_K}\left( t \right)} \right]^T}$, ${\bf{\tilde s}}\left( t \right) = {\left[ {{{\tilde s}_1}\left( t \right),\ldots ,{{\tilde s}_K}\left( t \right)} \right]^T}$, ${s_k}\left( t \right)= {s_k}\left( t \right)\sum\nolimits_{l = 1}^L {{\gamma _{k,l}}\left( t \right)}$, and ${{\tilde s}_k}\left( t \right) = {s_k}\left( t \right)\sum\nolimits_{l = 1}^L {{\gamma _{k,l}}\left( t \right){{\tilde \theta }_{k,l}}\left( t \right)}$.

\section{Two-Stage DOA Estimation}
In this section, we first exploit the output of PCAs based on  \eqref{4} to obtain initial DOA estimates. These estimates are then utilized to obtain the gain-phase uncertainties, which are finally compensated to achieve improved DOA estimation.
\subsection{First-Stage DOA Estimation}
Collecting $N$ snapshots, the matrix form of array output can be written as ${\bf{Z}}\left( t \right) = \left[ {{\bf{z}}\left( 1 \right),{\bf{z}}\left( 2 \right), \ldots ,{\bf{z}}\left( N \right)} \right]$. Based on the statistical expectation and sample average approximation, its covariance matrix can be calculated by
\begin{multline}\label{11}
\hspace{-5mm}{\bf{R}}   = {\bf{\Xi }}\left( {\bf{g}} \right){\bf{A}}{{\bf{R}}_s}{{\bf{A}}^H}{{\bf{\Xi }}^H}\left( {\bf{g}} \right) +{\bf{\Xi }}\left( {\bf{g}} \right){\bf{A'}}{{\bf{R}}_{\tilde s}}{{{\bf{A'}}}^H}{{\bf{\Xi }}^H}\left( {\bf{g}} \right) + \sigma _n^2{{\bf{I}}_M}  \\
\approx {1 \mathord{\left/ {\vphantom {1 N}} \right. \kern-\nulldelimiterspace} N}\sum\nolimits_{t = 1}^N {{\bf{z}}\left( t \right){{\bf{z}}^H}\left( t \right)}.\qquad\quad\qquad\quad\qquad\quad\quad\quad
\end{multline}
where ${{\bf{R}}s} = \mathbb{E}\left\{ s(t)s^H(t)\right\}$ is the source covariance matrix, and ${{\bf{R}}{\tilde s}} = \mathbb{E}\left\{ \tilde s(t)\tilde s^H(t)\right\}$ is the covariance matrix of the angular spread components. Further, it should be noted that the first row of ${{\bf{A'}}}$ (the derivative matrix of the steering vector) is a $1\times K$ all-zero vector. This property indicates that the first array element is immune to angular spread effects. Therefore, to eliminate   the impact of angular spread, we extract the first $M_c\times 1$ elements from ${\bf{R}}$ to form the covariance vector ${{\bf{r}}_c}$ (see  \eqref{12} below), which corresponds to the well-calibrated array elements. Assuming that the noise variance $\sigma_n^2$ is known $a$ $priori$ or pre-estimated via the maximum likelihood criterion \cite{ref16}, and subtracting the noise term from \eqref{11}, ${{\bf{r}}_c}$ is given by
\begin{equation}\label{12}
{{\bf{r}}_c} \approx {{\bf{A}}_c}{{\bf{R}}_s}{{\bf{1}}_{K \times 1}} = {{\bf{A}}_c}{\bf{p}},
\end{equation}
where ${\bf{p}} = {{\bf{R}}_s}{{\bf{1}}_{K \times 1}} = {\left[ {{p_1}, \ldots ,{p_K}} \right]^T}$ with ${p_k}$ being the power of ${{s_k}\left( t \right)}$, and ${\bf{A}}_c$ represents the first $M_c$ rows of $\bf{A}$. Since ${\bf{p}}$ is a positive vector, to further improve the resolution of DOA estimation, the conjugate symmetry strategy is adopted to construct the following augmented covariance vector as
\begin{equation}\label{13}
{{\bf{r}}_1} = {\left[ {{{\bf{\Pi }}_{{M_c}}}{\bf{r}}_c^ * ,{{\bf{r}}_c}\left( {2:{M_c}} \right)} \right]^T} \approx {\bf{Bp}},
\end{equation}
where ${\bf{B}} = \left[ {{\bf{b}}\left( {{\theta _1}} \right), \ldots ,{\bf{b}}\left( {{\theta _K}} \right)} \right]$ with its $k$-th column given by ${\bf{b}}\left( {{\theta _k}} \right) = {\left[ {{e^{j\left( {{M_c} - 1} \right)\pi \sin {\theta _k}}}, \ldots ,1, \ldots ,{e^{ - j\left( {{M_c} - 1} \right)\pi \sin {\theta _k}}}} \right]^T}$.

Let ${\bf{\Phi }}$ be the sparse representation matrix of ${\bf{B}}$, and ${{\bf{p}}_G}$ the $K$-sparsity vector with respect to $\bf{p}$. Then, initial DOA estimates can be obtained by solving the following $\ell_2$-$\ell_1$ norm minimization problem
\begin{equation}\label{14}
\begin{array}{l}
{P_1}:{{{\bf{\hat p}}}_G} = \arg \mathop {\min }\limits_{{{\bf{p}}_G}} \left\| {{{\bf{r}}_1} - {\bf{B}}{{\bf{p}}_G}} \right\|_2^2 + \lambda {\left\| {{{\bf{p}}_G}} \right\|_1},\\
{P_2}:\{ {{\hat \theta }_k}\} _{k = 1}^K  = \arg \mathop {\max }\limits_\theta  f_d\left( {{{{\bf{\hat p}}}_G}} \right),\theta  \in \left( { - {{90}^ \circ },{{90}^ \circ }} \right],
\end{array}
\end{equation}
where $f_d\left( {{{{\bf{\hat p}}}_G}} \right)$ returns the indices of nonzero elements in ${{{{\bf{\hat p}}}_G}}$, and $\lambda$ is the penalty parameter and selected here by the L-curve method.

 \subsection{Second-Stage DOA Estimation}
With the use of the  DOA estimates $\{ {{\hat \theta }_k}\} _{k = 1}^K$ , we can reconstruct the array manifold matrices $\hat{\bf{A}}$ and $\hat{{\bf{B}}}$ corresponding to these estimated angles. Here, ${\bf{\hat A}}= [ {{\bf{a}}( {{\hat \theta _1}} ),\ldots ,{\bf{a}}( {{\hat \theta _K}} )} ]$ contains the steering vectors for all $M$ array elements, while ${\bf{\hat B}} = [ {{\bf{b}}( {{\hat \theta _1}} ), \ldots ,{\bf{b}}( {{\hat \theta _K}})}]$ represents the augmented steering vectors using only the first $M_c$ calibrated elements. This allows us to obtain a refined power estimate
\begin{equation}\label{15}
{\bf{\hat p}} = {({{{\bf{\hat B}}}^H}{\bf{\hat B}})^{ - 1}}{{{\bf{\hat B}}}^H}{{\bf{r}}_1}.
\end{equation}
To estimate the gain-phase uncertainties, we extract the first column of ${\bf{R}}$ to form ${{\bf{r}}_2}$ which represents the correlation between all array elements and the first (calibrated) element:
\begin{equation}\label{16}
{{\bf{r}}_2} \approx {\bf{\Xi }}\left( {\bf{g}} \right){\bf{A}}{{\bf{R}}_s}{{\bf{1}}_{K \times 1}} = {\bf{\Xi }}\left( {\bf{g}} \right){\bf{Ap}} = {\bf{g}} \odot {\bf{Ap}}.
\end{equation}
The gain-phase uncertainty of each uncalibrated element can then be estimated by comparing the actual observation ${{\bf{r}}_2}$ with the ideal response $\bf{v}= {\bf{{\hat A}{\hat p}}}$:
\begin{equation}\label{17}
{{\hat g}_m} = {{\bf{r}}_2}\left( m \right)/{\bf{v}}\left( m \right),m \in [{M_c} + 1,M]
\end{equation}
where ${\bf{v}}\left( m \right)$ stands for the $m$-th element of ${\bf{{\hat A}{\hat p}}}$. Accordingly,  we  now utilize all $M$ array elements for improved DOA estimation. First, we reconstruct the gain-phase uncertainty  matrix ${\bf{\hat \Xi }}$ and compensate its effect in  ${{\bf{r}}_2}$:
\begin{equation}\label{18}
{{\bf{r}}_3} = {({{{\bf{\hat \Xi }}}^H}({\bf{\hat g}}){\bf{\hat \Xi }}({\bf{\hat g}}))^{ - 1}}{{{\bf{\hat \Xi }}}^H}({\bf{\hat g}}){{\bf{r}}_2} \approx {\bf{Ap}}.
\end{equation}
With the compensated array response ${{\bf{r}}_3}$, we can now apply the same conjugate symmetry strategy as in the first stage, but this time using all $M$ array elements instead of just $M_c$ calibrated ones:
\begin{equation}\label{19}
{{\bf{r}}_4} = {\left[ {{{\bf{\Pi }}_{{M}}}{\bf{r}}_3^ * ,{{\bf{r}}_3}\left( {2:{M}} \right)} \right]^T} \approx {\bf{Cp}}
\end{equation}
where ${\bf{C}} = \left[ {{\bf{c}}\left( {{\theta _1}} \right), \ldots ,{\bf{c}}\left( {{\theta _K}} \right)} \right]$ with its $k$-th column given by ${\bf{c}}\left( {{\theta _k}} \right) = {\left[ {{e^{j\left( {M - 1} \right)\pi \sin {\theta _k}}}, \ldots ,1, \ldots ,{e^{ - j\left( {M - 1} \right)\pi \sin {\theta _k}}}} \right]^T}$.

On the other hand, it is emphasized that the first-stage DOA estimation does not consider the impact of the finite number of snapshots and the model bias between the GAM and true array manifold. Taking both adverse factors into consideration, we employ the perturbation matrix ${\bf{\Gamma }}$ and rewrite ${{\bf{r}}_4}$ accurately in a sparse representation manner as ${{\bf{r}}_4} = \left( {{\bf{\Psi }} + {\bf{\Gamma }}} \right){{\bf{p}}_{\bar G}}$, where ${\bf{\Psi }}$ denotes the sparse representation of $\bf{C}$.

Next, the following sparse total least-squares minimization optimization problem is constructed to   achieve robust and refined DOA estimation:
\begin{equation}\label{20}
\begin{array}{l}
\{ {{{{\bf{\hat p}}}_{\bar G}},{\bf{\hat \Gamma }}} \} = \arg \mathop {\min }\limits_{{{\bf{p}}_{\bar G}},{\bf{\Gamma }}} \| {{\bf{\Gamma }}} \|_F^2 + \lambda {\left\| {{{\bf{p}}_{\bar G}}} \right\|_1}\\
\quad\quad s.\;t.\;\;{{\bf{r}}_4} = \left( {{\bf{\Psi }} + {\bf{\Gamma }}} \right){{\bf{p}}_{\bar G}}
\end{array}
\end{equation}
which can be relaxed as an unconstrained form as:
\begin{equation}\label{21}
\{ {{{{\bf{\hat p}}}_{\bar G}},{\bf{\hat \Gamma }}} \} = \arg \mathop {\min }\limits_{{{\bf{p}}_{\bar G}},{\bf{\Gamma }}} \left\| {{{\bf{r}}_4} - \left( {{\bf{\Psi }} + {\bf{\Gamma }}} \right){{\bf{p}}_{\bar G}}} \right\|_2^2 + \| {{\bf{\hat \Gamma }}} \|_F^2 + \lambda {\left\| {{{\bf{p}}_{\bar G}}} \right\|_1}.
\end{equation}

By exploiting the alternating descent algorithm,  \eqref{21} can be solved efficiently. Given ${{{\bf{\hat \Gamma }}}^{(i)}}$ at the $(i-1)$-th iteration, ${{{\bf{p}}_{\bar G}}}$ is updated at the $i$-th iteration as
\begin{equation}\label{22}
{\bf{\hat p}}_{\bar G}^{(i)} = \arg \mathop {\min }\limits_{{\bf{p}}_{\bar G}} \| {{{\bf{r}}_4} - ( {{\bf{\Psi }} + {{{\bf{\hat \Gamma }}}^{(i-1)}}} ){{\bf{p}}_{\bar G}}} \|_2^2 + \lambda {\left\| {{{\bf{p}}_{\bar G}}} \right\|_1}
\end{equation}
and with available ${\bf{\hat p}}_{\bar G}^{(i)}$, ${{{\bf{\hat \Gamma }}}^{(i)}}$ is updated as
\begin{equation}\label{23}
{{{\bf{\hat \Gamma }}}^{(i)}} = \arg \mathop {\min }\limits_{\bf{\Gamma }} \| {{{\bf{r}}_4} - ( {{\bf{\Psi }} + {\bf{\Gamma }}} ){\bf{\hat p}}_{\bar G}^{(i)}} \|_2^2 + \| {{\bf{\Gamma }}} \|_F^2.
\end{equation}

The optimization subproblems  \eqref{20} and  \eqref{22} are both convex, and therefore can be solved efficiently. The iteration is terminated when ${\| {{\bf{\hat p}}_{\bar G}^{(i)} - {\bf{\hat p}}_{\bar G}^{(i-1)}} \|_2} < \varepsilon $ or the user defined maximum number of iterations $I_t$ is reached. Finally, the improved DOA estimates $\{ {{\hat \theta }^{(f)}_k}\} _{k = 1}^K$ are computed by  identifying the positions of the $K$ non-zero elements in ${\bf{\hat p}}_{\bar G}^{(I_t)}$.


\emph{Remark 1:} It is well known that the sparse total least-squares techniques hold great robustness against both perturbations and noise. Meanwhile, the effective array aperture or the number of degrees of freedom (DOFs) associated with ${\bf{r}}_4$ is significantly increased in comparison with that of $\bf{r}_1$, implying that an improved DOA estimation accuracy is attained  at the second stage.

\begin{figure*}[t]
	\centering  
	\vspace{-0.35cm} 
	\subfigtopskip=2pt 
	\subfigbottomskip=-2pt 
	\subfigcapskip=-5pt 
	\subfigure[]{\includegraphics[width=2.4in]{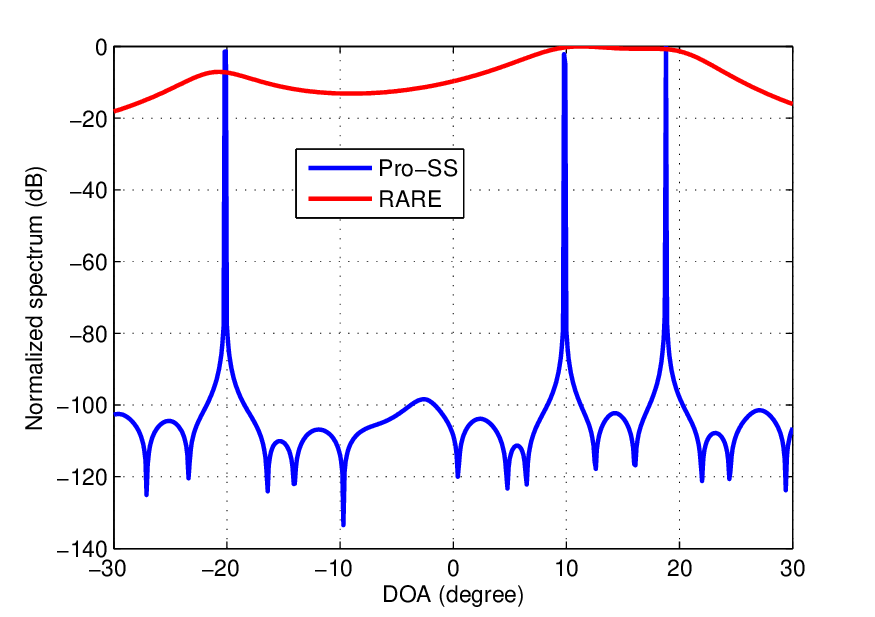}\label{fig1.a}}\hspace{-2mm}  
	\subfigure[]{\includegraphics[width=2.4in]{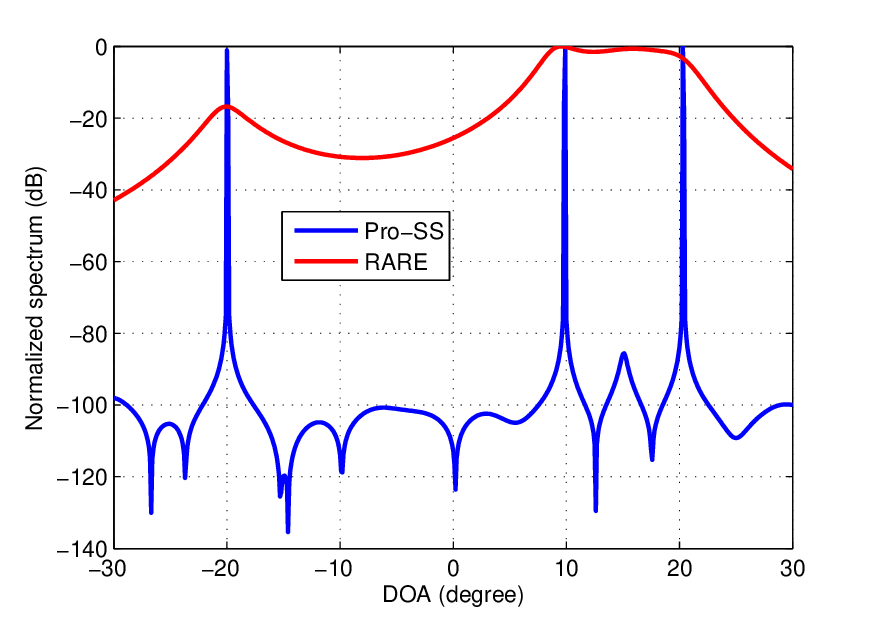}\label{fig1.b}}\hspace{-2mm}
	\subfigure[]{\includegraphics[width=2.4in]{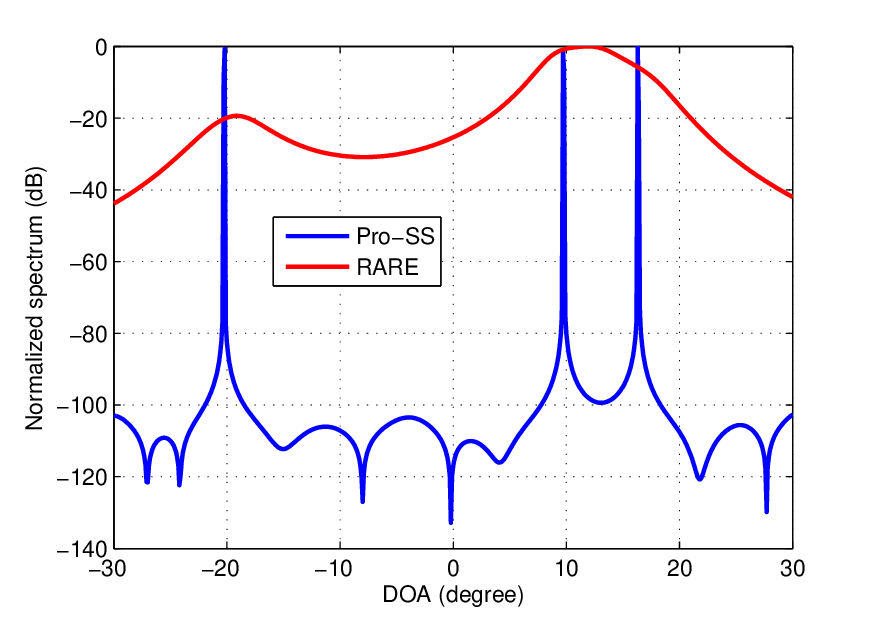}\label{fig1.c}}\hspace{-2mm}
	\caption{Normalized spectrum for Pro-SS and RARE under different settings: (a) ${\sigma _\theta=1.5^\circ}$, SNR=-6 dB, minimum angle separation = $10^\circ$; (b) ${\sigma _\theta=1.5^\circ}$, SNR=6 dB, minimum angle separation = $10^\circ$; (c) ${\sigma _\theta=2.5^\circ}$, SNR=6 dB, minimum angle separation = $6^\circ$.}
	\label{level}
\end{figure*}
\begin{figure*}[t]
	\centering  
	\vspace{-0.35cm} 
	\subfigtopskip=2pt 
	\subfigbottomskip=-2pt 
	\subfigcapskip=-5pt 
	\subfigure[]{\includegraphics[width=2.4in]{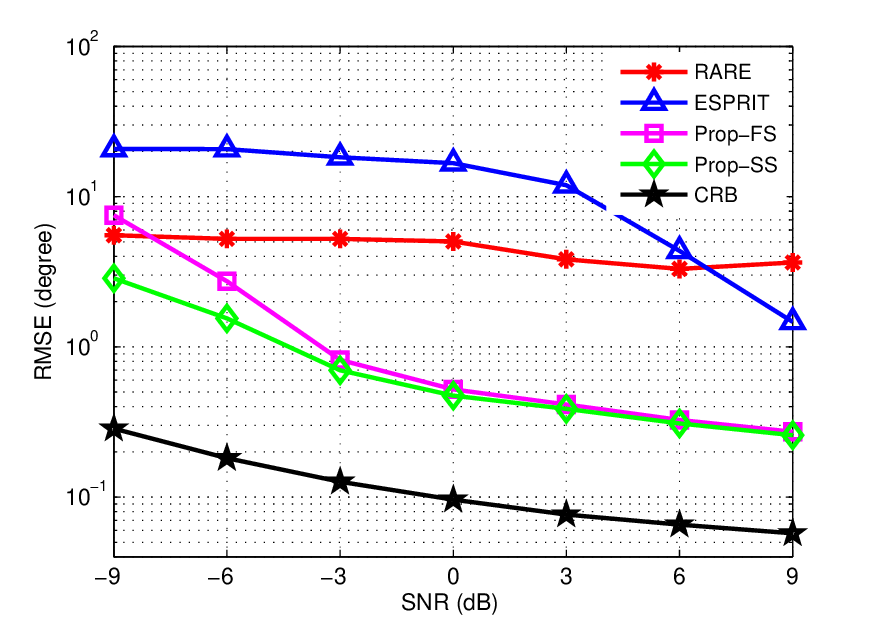}\label{fig2.a}}\hspace{-2mm}  
	\subfigure[]{\includegraphics[width=2.4in]{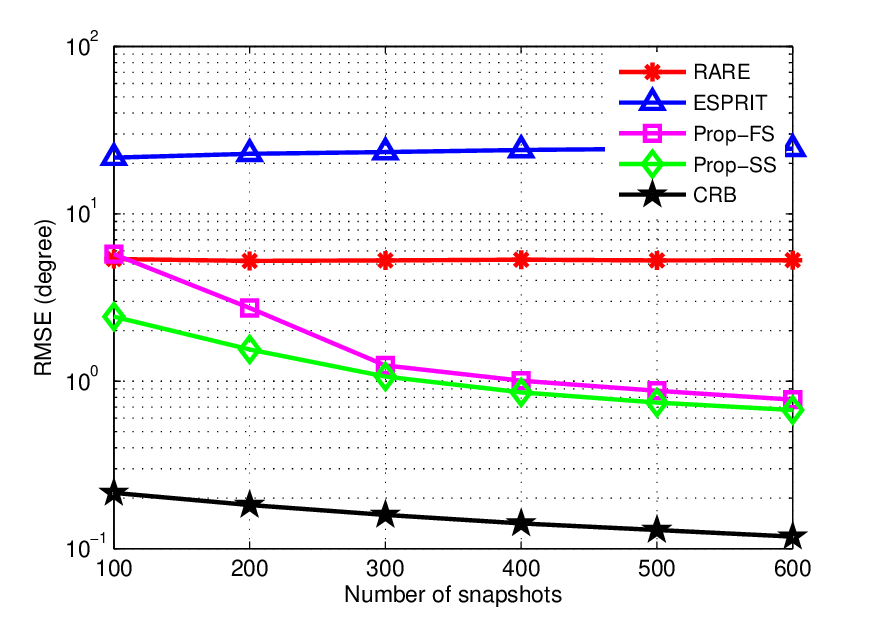}\label{fig2.b}}\hspace{-2mm}
	\subfigure[]{\includegraphics[width=2.4in]{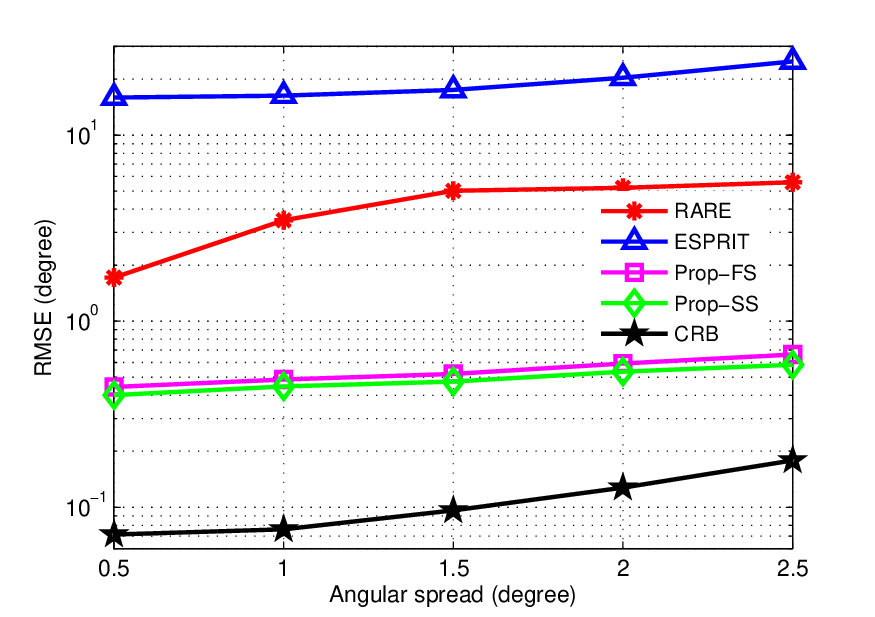}\label{fig2.c}}\hspace{-2mm}
	\caption{RMSE of DOA estimates for different methods under different settings: (a) RMSE versus SNR; (b) RMSE versus   number of snapshots; (c) RMSE versus angular spread.}
	\label{level}
\end{figure*}
\section{Simulation Results}
In this section, the performance of the two-stage DOA estimation method is assessed, and compared with that of the RARE in \cite{ref8}, ESPRIT in \cite{ref13} and Cram\'{e}r-Rao Bound (CRB) in \cite{ref13}, where the first-stage and second-stage estimations are respectively designated as Pro-FS and Pro-SS. It should be noted that the RARE based method is not suitable for dealing with array gain-phase uncertainties, and thus we utilize the gain-phase uncertainty estimation result obtained from our method to compensate it for comparison. In the simulations, a ULA with $M=16$ sensors is employed and the root mean square error (RMSE) obtained by averaging 300 independent Monte-Carlo trials is selected to evaluate the performance of different methods. The gain and phase uncertainties are respectively modeled as ${\rho _m} = 1 + \sqrt {12} {\sigma _\rho }{\eta_m},\;{\phi _m} = \sqrt {12} {\sigma _\phi }{\mu_m}$,
where ${\eta_m}$ and ${\mu_m}$ are uniformly i.i.d. random variables within $[-0.5,0.5]$, and ${\sigma _\rho }$ and ${\sigma _\phi }$ represent the standard deviations of ${\eta_m}$ and ${\mu_m}$, respectively.

In the first simulation, the normalized spatial spectra of the proposed method and RARE for three ID sources are plotted in Fig. 1. The parameters of three ID sources   $\left({\theta _1},{\theta _2},{\theta _3} \right) = \left( {-20}^ \circ,{10}^ \circ ,{20}^ \circ\right)$, ${\sigma_\theta=\{\sigma_{\theta_k}\}_{k=1}^3=1.5^\circ}$, with $N=200$ and SNR=-6 dB in Fig. 1(a), from which we   observe that the proposed method   resolve three sources well, albeit with some estimation bias, while the compared RARE fails. In Fig. 1(b), SNR is increased to 6 dB, and we  see that although RARE  distinguish two closely-spaced sources, its spectral peaks are much less pronounced. In contrast, our solution  distinguish them very clearly. In Fig. 1(c), we fix SNR at 6 dB, while setting  the minimum angle separation to $6^\circ$ and   ${\sigma_\theta=2.5^\circ}$. Under such a smaller angle separation scenario,  the proposed method  still resolve two closely-spaced ID sources successfully, demonstrating a high resolution performance.

In the second simulation, the performance of different methods is studied with respect to SNR, number of snapshots $N$, as well as angular spread $\sigma_{\theta}$, with $M_c=8$, ${\sigma _\rho }=0.1$ and ${\sigma _\phi }=40^ \circ$. The results are shown in Fig. 2, where in both Figs. 2(a) and 2(b), the settings are $\left( {{\theta _1},{\sigma _{{\theta _1}}},{\theta _2},{\sigma _{{\theta _2}}}} \right) = \left( {{{10}^ \circ },{{1.5}^ \circ },{{20}^ \circ },{{1.5}^ \circ }} \right)$, while $N$ is fixed at 200, and SNR varies from -9 dB to 9 dB in Fig. 2(a), and SNR is fixed at -6 dB, and $N$ varies from 100 to 600 in Fig. 2(b). It is observed that the proposed method   provide an improved performance compared with the remaining algorithms. Meanwhile, due to the increase of DOFs and robustness to perturbations of the sparse total least-squares approach adopted at the second stage, Pro-SS performs better than Pro-FS in the whole observed regions, especially under low SNRs and small number of snapshots, showing the effectiveness and superiority of the   two-stage DOA estimator. In Fig. 2(c), SNR and $N$ are respectively set to 0 dB and 200, two DOAs remain unchanged, but the angular spread $\sigma_{\theta}$ for both ID sources varies from $0.5^ \circ$ to $2.5^ \circ$. It is observed that the performance of all methods gets worse with the increase of angular spread. However, the proposed solution has the least degree of performance degradation.

\section{Conclusion}
In this paper, we have presented a two-stage sparse DOA estimation framework for ID sources under gain-phase uncertainties.  The first stage aims to obtain initial DOA estimates using CPAs, while the second stage has compensated for gain-phase uncertainties and has refined the estimation through sparse total least-squares optimization. Simulation results have demonstrated the superiority over  existing approaches in terms of accuracy and robustness.

\end{document}